\documentclass[12pt,a4paper]{article}
\usepackage{amsmath,amssymb}

\begin{document}

\title{Particle Spectrum in Modified NMSSM}

\author{R.B.Nevzorov, K.A.Ter--Martirosyan, and M.A.Trusov}

\maketitle

\begin{abstract}

\noindent The restrictions on the lightest Higgs boson mass in the
minimal supersymmetric models are briefly reviewed. The particle
spectrum is considered in the framework of the simplest
modification of NMSSM that allows to avoid the domain wall problem
and to get the self--consistent solution in the strong Yukawa
coupling limit. The lightest Higgs boson mass in the investigated
model can reach $125\text{~GeV}$ at values of $\tan\beta\ge 1.9$
and does not exceed $130.5\pm 3.5\text{~GeV}$.

\end{abstract}

\newpage

An important feature of supersymmetric models is the existence of
a light Higgs boson in the CP--even Higgs sector. In the Minimal
Supersymmetric Standard Model (MSSM) at the tree level the
lightest Higgs boson mass does not exceed the $Z$--boson mass, but
the loop corrections from the $t$--quark and its superpartners
significantly raise the upper bound on $m_h$:
\[
m_h^2\le M_Z^2\cos^2 2\beta+\Delta,
\]
where $\Delta$ are the loop corrections.

In the case of moderate values of $\tan\beta$ a considerable
fraction of the solutions of the system of MSSM renormalization
group equations is focused near the infrared quasi--fixed point,
which allows one to get rather strong constraint on the lightest
Higgs boson mass \cite{1}: $m_h\le 94\pm 5\text{~GeV}$. This
conflicts with an experimental restriction from LEP\,II: $m_h\ge
113.3\text{~GeV}$. So, this scenario is excluded nowadays. With
increasing of $\tan\beta$ the upper bound on $m_h$ grows, and for
$\tan\beta\gg 1$ reaches $125-128\text{~GeV}$. But this scenario
is also objectionable for two reasons. The first problem is the
too fast proton decay in SUSY GUTs. For the major decay mode $p\to
\bar{\nu}K^{+}$ the partial life--time $\tau_p$ is inversely
proportional to $\tan^2\beta$. When $\tan\beta$ is large enough
the theoretical prediction for $\tau_p$ becomes larger than the
corresponding experimental restriction. The second problem
concerns flavour--changing neutral currents. The branching ratio
$b\to s\gamma$ rises with $\tan\beta$ as $Br(b\to s\gamma)\sim
\tan^2\beta$. Thus the values of $\tan\beta\gtrsim 40$ lead to the
unacceptable large flavour--changing transitions. All this
stimulates investigations of the Higgs sector in nonminimal
supersymmetric models.

The simplest extension of the MSSM which can conserve the
unification of the gauge constants and raise the upper bound on
the mass of the lightest Higgs boson is the Next--to--Minimal
Supersymmetric Standard Model (NMSSM) \cite{2}. In addition to
doublets $H_1$ and $H_2$, the Higgs sector of this model contains
an additional singlet superfield $Y$. The superpotential of the
Higgs sector of the NMSSM includes two terms:
$\lambda\hat{Y}(\hat{H}_1\hat{H}_2)+\frac{\varkappa}{3}\hat{Y}^3$.
It is invariant under $Z_3$ group discrete transformations. After
electroweak symmetry breaking the singlet field $Y$ acquires a
nonzero vacuum expectation value ($\langle Y\rangle=y/\sqrt{2}$),
so the effective $\mu$--term is generated.

The upper bound on the mass of the lightest Higgs boson in the
NMSSM is calculated by formula:
\[
m_h^2\le\frac{\lambda^2}{2}v^2\sin^2 2\beta+M_Z^2\cos^2
2\beta+\Delta.
\]
This upper limit considerably exceeds the corresponding bound in
the MSSM only in the strong Yukawa coupling limit when the Yukawa
constants on the Grand Unification scale $M_X$ are substantially
larger than the gauge constant $g_{\text{GUT}}$.

Unfortunately, in this regime for a minimal choice of fundamental
parameters it is impossible to get a self--consistent solution of
the equations defining the classical minimum of the potential of
the model. Such solutions appear only for very small values of
$\lambda$ and $\varkappa$ when the upper bound on $m_h$ is the
same as in the MSSM. Moreover, due to the $Z_3$ symmetry three
degenerate vacuum configurations arise, that leads to the domain
wall problem \cite{3}.

Thus, in order to avoid the domain wall problem and get the
self--consistent solution in the strong Yukawa coupling limit, one
has to modify the NMSSM. The simplest way is to introduce the
bilinear terms $\mu(\hat{H}_1\hat{H}_2)$ and $\mu'\hat{Y}^2$ in
the superpotential which are not forbidden by electroweak
symmetry. At the same time one can omit the coupling $\varkappa$,
that allows to simplify the analysis of the modified NMSSM. For
$\varkappa=0$ the upper bound on the lightest Higgs boson mass
reaches its maximum value. Neglecting all the Yukawa constants
except for $h_t$ and $\lambda$ one gets the following expression
for the modified NMSSM (MNSSM) superpotential \cite{4}:
\[
W_{\text{MNSSM}}=\mu(\hat{H}_1\hat{H}_2)+\mu'\hat{Y}^2+\lambda
\hat{Y}(\hat{H}_1\hat{H}_2)+ h_t(\hat{H}_2\hat{Q})\hat{U}^C_R.
\]
The bilinear terms in the superpotential break the $Z_3$ symmetry
and the domain walls do not arise, because the degenerated vacua
do not exist. The introduction of the parameter $\mu$ permits to
obtain the self--consistent solution of algebraic equations in the
strong Yukawa coupling limit.

The $\mu$--terms in the $W_{\text{MNSSM}}$ lead to the appearance
of bilinear scalar couplings $B$ and $B'$ in the effective
potential of the Higgs fields. They arise as a result of the soft
supersymmetry breaking. For a minimal choice of the fundamental
parameters all soft scalar masses, all trilinear couplings, and
all bilinear couplings have to be put equal at the GUT scale
$M_X$. Thus in addition to the parameters of the SM the MNSSM
contains seven independent ones:
\[ \lambda,\mu,\mu',A,B_0,m_0^2,M_{1/2}. \]
Although the parameter space of MNSSM is enlarged the theory does
not lose the predictive capacity. In the strong Yukawa coupling
limit the solutions of renormalization group equations are
concentrated near the quasi fixed point \cite{5}, that corresponds
to the value of $\tan\beta\approx 1.9$. Near the quasi fixed point
the number of degrees of freedom of MNSSM parameter space
diminishes.

The Higgs sector of the MNSSM includes three CP--even, two CP--odd
and one charged massive state. There are two parts of the
parameter space where the self--consistent solution exists. In one
of them the parameters $\mu$ and $\mu'$ have the same signs. The
mass of the lightest CP--even Higgs boson in this region is always
smaller than the upper bound on $m_h$ in the MSSM. So, it is
excluded now by LEP\,II data. In the other region, where the signs
of $\mu$ and $\mu'$ are opposite, the value of $m_h$ is larger
than in the MSSM. It allows to relax the strong lower bound on
$\tan\beta$ which comes from the analysis of the MSSM Higgs
sector. The qualitative pattern of the particle spectrum in the
considered part of MNSSM parameter space is sensitive mostly to
the choice of two parameters: $\mu'$ and the value of the SUSY
breaking scale $M_S$. For the case of heavy supersymmetric
particles ($M_S\gg M_Z$) the $(3\times 3)$ mass matrix of the
CP--even Higgs sector has the hierarchical structure. Considering
$(M_Z/M_S)^2$ as a small parameter one can diagonalize it using
the method of perturbation theory \cite{6}. The same method can be
applied for the diagonalization of $(5\times 5)$ neutralino mass
matrix.

The heaviest particle in the modified NMSSM is the CP--even Higgs
boson corresponding to the neutral field $Y$. Its mass $m_S$ is
proportional to $\mu'$ and substantially larger than the scale of
supersymmetry breaking. The mass of the other heavy CP--even Higgs
boson $m_H$ is almost insensitive to the value of $\mu'$, while
the masses of the CP--odd states rise with increasing of $\mu'$.
The heaviest fermion in the MNSSM is the neutralino
$\tilde{\chi}_5$ which is the superpartner of the singlet field
$Y$. Its mass $m_{\tilde{\chi_5}}$ is proportional to $\mu'$ too.
The limit $\mu'\gg M_S$, when neutralino $\tilde{\chi_5}$, the
CP--even, and CP--odd scalar singlet fields become very heavy,
corresponds to the minimal SUSY model.

For a reasonable set of the parameters of MNSSM gluinos, squarks,
heavy CP--even and CP--odd Higgs bosons are substantially heavier
than sleptons, lightest charginos and neutralinos. The only
exception is one of the CP--odd Higgs bosons whose mass decreases
if $\mu'$ diminishes. However, even if its mass is low enough, for
example of the order of $M_Z$, it will be quite difficult to
observe it in future experiments because the main contribution to
its wave function is given by the CP--odd component of the singlet
field $Y$.

The lightest CP--even Higgs boson mass in the modified NMSSM may
reach $127\text{~GeV}$ even for the comparatively low value of
$\tan\beta\approx 1.9$. For each value of $\tan\beta$ one can
choose the value of $\mu'$ so that $m_h$ reaches its theoretical
upper bound. The obtained function $m_h(\tan\beta)$ has a maximum
for $\tan\beta\approx 2.2\div 2.4$ which corresponds to the strong
Yukawa coupling limit. The numerical analysis \cite{4} reveals
that the mass of the lightest CP--even Higgs boson in the MNSSM is
always smaller than $130.5\pm 3.5\text{~GeV}$, where the
uncertainty is mainly due to the error in the top quark mass.

\section*{Acknowledgments}

The authors are grateful to D. I. Kazakov, L. B. Okun, and M. I.
Vysotsky for stimulating discussions. Our work was supported by
the Russian Foundation for Basic Research (RFBR) (projects \#\#
00-15-96786 and 00-15-96562).


\begin{thebibliography}{9}

\bibitem{1} J. A. Casas, J. R. Espinosa, and
H. E. Haber, Nucl. Phys. B {\bfseries 526}, 3 (1998); S. A. Abel
and B. C. Allanach, Phys. Lett. B {\bfseries 431}, 339 (1998); G.
K. Yeghiyan, M. Jurcisin, and D. I. Kazakov, Mod. Phys. Lett. A
{\bfseries 14}, 601 (1999).

\bibitem{2} P. Fayet, Nucl. Phys. B {\bfseries 90}, 104 (1975);
M. I. Vysotsky and K. A. Ter--Martirosyan, Sov. Phys. JETP
{\bfseries 63}, 489 (1986).

\bibitem{3} S. A. Abel, S. Sarkar, and P. L. White, Nucl. Phys.
B {\bfseries 454}, 663 (1995).

\bibitem{4} R. B. Nevzorov and M. A. Trusov, Sov. Phys. JETP
{\bfseries 91}, 1079 (2000).

\bibitem{5} R. B. Nevzorov and M. A. Trusov, Phys. Atom.
Nucl. {\bfseries 64}, 1299 (2001); Phys. Atom. Nucl. {\bfseries
64}, 1589 (2001).

\bibitem{6} P. A. Kovalenko, R. B. Nevzorov, and K. A. Ter--Martirosyan,
Phys. Atom. Nucl. {\bfseries 61}, 812(1998).

\end{thebibliography}
\end{document}